\documentclass[twocolumn,showpacs,amsmath,amssymb,aps,prd]{revtex4}

\usepackage{amsmath}

\newcommand{\itGamma}{{\mathit{\Gamma}}}
\newcommand{\del}{\partial}

\newcommand{\cK}{{\cal K}}
\newcommand{\cM}{{\cal M}}

\newcommand{\cR}{{\cal R}}
\newcommand{\cB}{{\cal B}}
\newcommand{\cT}{{\cal T}}
\newcommand{\ds}{\displaystyle}

\newcommand{\lc}{\varepsilon}

\newcommand{\diag}{\mathop{\rm diag}\nolimits}
\newcommand{\orto}{{\scriptscriptstyle\perp}}

\newcommand{\Pn}{ {P_{\orto}}\vphantom{P} }
\newcommand{\cfl}[2]{{\textstyle {{#1}\brace {#2}}}}

\newcommand{\tm}{ {\bar{m}}\vphantom{m} }
\newcommand{\Ka}{\!\stackrel{\scriptscriptstyle\rm A}{K}\vphantom{K}\!\!}
\newcommand{\rmd}{d}

\begin{document}

\title{Test membranes in Riemann-Cartan spacetimes}

\author{Milovan Vasili\'c}
 \email{mvasilic@ipb.ac.rs}
\author{Marko Vojinovi\'c}
 \email{vmarko@ipb.ac.rs}
\affiliation{Institute of Physics, P.O.Box 57, 11001 Belgrade, Serbia}

\date{\today}

\begin{abstract}
The dynamics of brane-like extended objects in spacetimes with torsion is
derived from the conservation equations of stress-energy and spin tensors.
Thus obtained world-sheet equations are applied to macroscopic test
membranes made of spinning matter. Specifically, we consider membranes with
maximally symmetric distribution of stress-energy and spin. These are
characterized by two constants only: the tension and spin magnitude. By
solving the world-sheet equations, we discover a similarity between such
membranes in Riemann-Cartan backgrounds, and string theory membranes in
low-energy string backgrounds. In the second part of the paper, we apply
this result to cylindrical membranes wrapped around the extra compact
dimension of a $(D+1)$-dimensional spacetime. In the narrow membrane limit,
we discover how effective macroscopic strings couple to torsion. An observed
similarity with the string sigma model is noted.
\end{abstract}


\maketitle

\section{Introduction}\label{Sec1}

The problem of motion of brane-like extended objects in backgrounds of
nontrivial geometry is addressed by using some form of the
Mathisson-Papapetrou method \cite{1,2}. One starts with the covariant
conservation law of the stress-energy and spin tensors of matter fields, and
analyzes it under the assumption that matter is localized to resemble a
brane. In the lowest, single-pole approximation, the moving matter is viewed
as an infinitely thin brane. In the pole-dipole approximation, its non-zero
thickness is taken into account.

The known results concerning extended objects in Riemann-Cartan geometry can
be summarized as follows. Spinless particles in the single-pole
approximation obey the geodesic equation. In the pole-dipole approximation,
the rotational angular momentum of the localized matter couples to spacetime
curvature, and produces geodesic deviation \cite{1,2,3,4,5}. If particles
have spin, the curvature couples to the total angular momentum, and the
torsion to the spin alone \cite{6,7,8,9,10}. Higher branes have been
considered in \cite{11,12,12a}. In the spinless case, the curvature couples
to the internal angular momentum of a thick brane. The coupling disappears
in the limit of an infinitely thin brane \cite{11,12}. The dynamics of
spinning branes has been investigated in \cite{12a}. The spin-torsion
coupling has been derived for an arbitrary distribution of spinning matter
over the brane.

In this paper, we want to examine infinitely thin branes with maximally
symmetric distribution of stress-energy and spin. In the spinless case, such
branes are characterized by the tension alone, and are known as the
Nambu-Goto branes \cite{13,14}. These branes do not couple to torsion, and
our idea is to consider their minimal extension characterized by an
additional constant---the spin magnitude. The analysis along these lines has
already been attempted in \cite{12a}. The idea has been to try and find the
circumstances under which the classical dynamics of strings could possibly
resemble the dynamics of \cite{15,16,17,18,t7,t19}. Two major results have
been obtained. First, the nontrivial string-torsion coupling exists only if
the string is made of spinning matter. Second, if the torsion is totally
antisymmetric, the form of the obtained dynamics necessarily differs from
\cite{15,16,17,18,t7,t19}.

In this paper, we continue investigating the behaviour of brane-like
extended objects in Riemann-Cartan spacetime. We emphasize that our work is
not a part of the mainstream string theory considerations. Our brane-like
objects are made of conventional matter, and are used to probe
Riemann-Cartan geometry. The only connection with string theory is seen in
the form of our resulting equations. Namely, we notice a similarity between
the motion of macroscopic test membranes in Riemann-Cartan backgrounds and
the motion of string theory membranes in the low-energy string backgrounds.
Whether this is just a coincidence, or there is more content in this analogy
is not the subject of this paper.

In the subsequent paragraphs, we shall investigate the influence of torsion
on the dynamics of membrane-like extended objects. Our effort is motivated
by the observation that problems encountered in treating strings in
Riemann-Cartan spacetime are nicely solved if membranes are considered
instead. In particular, the nontrivial projection of the totally
antisymmetric spin tensor on the $p$-brane world-sheet is shown to exist
only if $p\geq 2$. This form of spin tensor proves to be the basic
ingredient in the construction of membranes with maximally symmetric
distribution of spin. We apply the $p$-brane equations of \cite{12a} to a
membrane characterized by two constants only: the tension and spin
magnitude. As a result, the behaviour of such membranes in Riemann-Cartan
backgrounds is shown to follow from an action functional. A resemblance with
the $\sigma$-model action of \cite{22} is noted. The latter describes a
membrane coupled to the string theory 3-form field.

The effective string dynamics is obtained in the narrow membrane limit. We
consider cylindrical membranes wrapped around the extra compact dimension of
a $(D+1)$-dimensional spacetime, and perform a $D+1\to D$ dimensional
reduction. As a result, a $\sigma$-model action similar to that of
\cite{15,16,17,18,t7,t19} is revealed.

The layout of the paper is as follows. In section \ref{Sec2}, we review the
basic notions of the multipole formalism developed in \cite{11,12,12a}. The
manifestly covariant world-sheet equations and boundary conditions are
explicitly displayed. In section \ref{Sec3}, the $p=2$ brane is
investigated. The free coefficients of the theory are chosen to define a
Nambu-Goto membrane with totally antisymmetric spin tensor parallel to the
world-sheet. This membrane is characterized by the constant tension and
spin, and its dynamics in Riemann-Cartan geometry is shown to resemble the
dynamics of elementary membranes in the low-energy string backgrounds
\cite{22}. In section \ref{Sec4}, a narrow membrane wrapped around the extra
compact dimension of a $(D+1)$-dimensional spacetime is considered. An
effective string dynamics is revealed in the limit of the small extra
dimension. Section \ref{Sec5} is devoted to concluding remarks.

Our conventions are the same as in  \cite{12}. Greek indices
$\mu,\nu,\dots$ are the spacetime indices, and run over $0,1,\dots,D-1$.
Latin indices $a,b,\dots$ are the world-sheet indices and run over
$0,1,\dots,p$. Latin indices $i,j,\dots$ refer to the world-sheet boundary
and take values $0,1,\dots,p-1$. The coordinates of spacetime, world-sheet
and world-sheet boundary are denoted $x^{\mu}$, $\xi^a$ and $\lambda^i$
respectively. The spacetime metric is denoted by $g_{\mu\nu}(x)$, and the
induced world-sheet metric by $\gamma_{ab}(\xi)$. The signature convention
is defined by $\diag(-,+,\dots,+)$, and the indices are raised using the
inverse metrics $g^{\mu\nu}$ and $\gamma^{ab}$.

\section{Multipole formalism}\label{Sec2}

It has been shown in \cite{11,12} that an exponentially decreasing
function can be expanded as a series of $\delta$-function derivatives. For
example, a tensor-valued function $F^{\mu\nu}(x)$, well localized around the
$p+1$-dimensional surface $\cM$ in $D$-dimensional spacetime, can be
decomposed in a manifestly covariant way as
\begin{equation} \label{jna1}
\begin{array}{ccl}
F^{\mu\nu}(x) & = & \ds \int_{\cM} d^{p+1}\xi \sqrt{-\gamma} \left[ M^{\mu\nu}
\frac{\delta^{(D)}(x-z)}{\sqrt{-g}} - \right. \\
 & & \ds \rule{0pt}{20pt} \left. - \nabla_{\rho} \left( M^{\mu\nu\rho}
\frac{\delta^{(D)}(x-z)}{\sqrt{-g}} \right) + \dots \right] . \\
\end{array}
\end{equation}
The surface $\cM$ is defined by the equation $x^{\mu}=z^{\mu}(\xi)$, where
$\xi^a$ are the surface coordinates, and the coefficients $M^{\mu\nu}(\xi)$,
$M^{\mu\nu\rho}(\xi), \dots$ are spacetime tensors called multipole
coefficients. Here, and throughout the paper, we make use of the surface
coordinate vectors
$$
u_a^{\mu} \equiv \frac{\del z^{\mu}}{\del\xi^a},
$$
and the surface induced metric tensor
$$
\gamma_{ab} = g_{\mu\nu} u_a^{\mu} u_b^{\nu}.
$$
The induced metric is assumed to be nondegenerate, $\gamma \equiv
\det(\gamma_{ab}) \neq 0$, and of Minkowski signature. The same holds for
the target space metric $g_{\mu\nu}(x)$ and its determinant $g(x)$. The
covariant derivative $\nabla_{\rho}$ is defined by the Levi-Civita
connection.

It has been shown in  \cite{12} that one may truncate the series in a
covariant way in order to approximate the description of matter. Truncation
after the leading term is called {\it single-pole} approximation, truncation
after the second term is called {\it pole-dipole} approximation. In the
single-pole approximation, one assumes that the brane has no thickness,
which means that matter is localized on a surface. All higher
approximations, including pole-dipole, allow for a nonzero thickness, and
thus, for the transverse internal motion.

The decomposition (\ref{jna1}) is particularly useful for the description of
brane-like objects in spacetimes of general geometry. In  \cite{12a}, the
pole-dipole approximation has been used to model the fundamental matter
currents --- stress-energy tensor $\tau^{\mu}{}_{\nu}$, and spin tensor
$\sigma^{\lambda}{}_{\mu\nu}$. The brane dynamics in Riemann-Cartan
backgrounds is obtained from the covariant conservation laws
\begin{subequations} \label{jna2}
\begin{equation} \label{jna2a}
\left( D_{\nu} + \cT^{\lambda}{}_{\nu\lambda} \right) \tau^{\nu}{}_{\mu} =
\tau^{\nu}{}_{\rho}\cT^{\rho}{}_{\mu\nu} +
\frac{1}{2}\sigma^{\nu\rho\sigma}\cR_{\rho\sigma\mu\nu},
\end{equation}
\begin{equation} \label{jna2b}
\left( D_{\nu} + \cT^{\lambda}{}_{\nu\lambda} \right)
\sigma^{\nu}{}_{\rho\sigma} = \tau_{\rho\sigma}-\tau_{\sigma\rho}.
\end{equation}
\end{subequations}
Here, $D_{\nu}$ is the covariant derivative with the nonsymmetric connection
$\itGamma^{\lambda}{}_{\mu\nu}$, which acts on a vector $v^{\mu}$ according
to the rule $D_{\nu}v^{\mu} \equiv \del_{\nu}v^{\mu} +
\itGamma^{\mu}{}_{\lambda\nu}v^{\lambda}$. The torsion
$\cT^{\lambda}{}_{\mu\nu}$ and curvature $\cR^{\mu}{}_{\nu\rho\sigma}$ are
defined in the standard way:
$$
\begin{array}{lcl}
\cT^{\lambda}{}_{\mu\nu} & \equiv & \itGamma^{\lambda}{}_{\nu\mu} -
\itGamma^{\lambda}{}_{\mu\nu}, \\
\cR^{\mu}{}_{\nu\rho\sigma} & \equiv &
\del_{\rho}\itGamma^{\mu}{}_{\nu\sigma} \! -
\del_{\sigma}\itGamma^{\mu}{}_{\nu\rho} +
\itGamma^{\mu}{}_{\lambda\rho}\itGamma^{\lambda}{}_{\nu\sigma} \! -
\itGamma^{\mu}{}_{\lambda\sigma}\itGamma^{\lambda}{}_{\nu\rho} . \\
\end{array}
$$
The covariant derivative $D_{\nu}$ is assumed to satisfy the metricity
condition $D_{\lambda}g_{\mu\nu}=0$. As a consequence, the connection
$\itGamma^{\lambda}{}_{\mu\nu}$ is split into the Levi-Civita connection
$\cfl{\lambda}{\mu\nu}$ and the contorsion $K^{\lambda}{}_{\mu\nu}$:
$$
\begin{array}{lcl}
\itGamma^{\lambda}{}_{\mu\nu} & = & \ds \cfl{\lambda}{\mu\nu} +
K^{\lambda}{}_{\mu\nu}, \\
K^{\lambda}{}_{\mu\nu} & \equiv & \ds -\frac{1}{2} \left(
\cT^{\lambda}{}_{\mu\nu} - \cT_{\nu}{}^{\lambda}{}_{\mu} +
\cT_{\mu\nu}{}^{\lambda} \right) . \\
\end{array}
$$
The curvature tensor can then be rewritten in terms of the Riemann curvature
$R^{\mu}{}_{\nu\rho\sigma} \equiv
\cR^{\mu}{}_{\nu\rho\sigma}(\itGamma\to\{\})$, and the Riemannian covariant
derivative $\nabla_{\mu} \equiv D_{\mu}(\itGamma\to\{\})$:
$$
\cR^{\mu}{}_{\nu\lambda\rho} = R^{\mu}{}_{\nu\lambda\rho} + 2
\nabla_{[\lambda}K^{\mu}{}_{\nu\rho]} + 2 K^{\mu}{}_{\sigma[\lambda}
K^{\sigma}{}_{\nu\rho]}.
$$

Given the system of conservation equations (\ref{jna2}),
one finds that the second one has no dynamical content. Indeed, the
antisymmetric part of the stress-energy tensor is completely determined by
the spin tensor. One can use (\ref{jna2b}) to eliminate $\tau^{[\mu\nu]}$
from (\ref{jna2a}), and thus obtain the conservation equations in which
only $\tau^{(\mu\nu)}$ and $\sigma^{\lambda\mu\nu}$ appear. The resulting
equation has the form
\begin{equation} \label{jna3}
\begin{array}{c}
\ds \nabla_{\nu} \left( \theta^{\mu\nu} - K^{[\mu}{}_{\lambda\rho}
\sigma^{\rho\lambda\nu]} - \frac{1}{2}
K_{\lambda\rho}{}^{[\mu}\sigma^{\nu]\rho\lambda} \right) = \\
\ds
= \frac{1}{2} \sigma_{\nu\rho\lambda} \nabla^{\mu} K^{\rho\lambda\nu}, \\
\end{array}
\end{equation}
where $\theta^{\mu\nu}=\theta^{\nu\mu}$ stands for the generalized
Belinfante tensor:
\begin{equation} \label{jna4}
\theta^{\mu\nu} \equiv \tau^{(\mu\nu)} - \nabla_{\rho}\sigma^{(\mu\nu)\rho}
- \frac{1}{2}K_{\lambda\rho}{}^{(\mu}\sigma^{\nu)\rho\lambda}.
\end{equation}
The independent variables $\theta^{\mu\nu}$ and $\sigma^{\mu\nu\rho}$ are in
$1$--$1$ correspondence with the original variables. The conservation law
in the form (\ref{jna3}) is the starting point for the derivation of brane
world-sheet equations.

In this paper, we are interested in infinitely thin branes, and therefore,
restrict our analysis to the single-pole approximation. The multipole
expansion of our basic variables then reads:
\begin{subequations} \label{jna5}
\begin{equation} \label{jna5a}
\theta^{\mu\nu} = \int_{\cM} \rmd^{p+1}\xi \sqrt{-\gamma} B^{\mu\nu}
\frac{\delta^{(D)}(x-z)}{\sqrt{-g}} ,
\end{equation}
\begin{equation} \label{jna5b}
\sigma^{\lambda\mu\nu} = \int_{\cM} \rmd^{p+1}\xi \sqrt{-\gamma}
C^{\lambda\mu\nu} \frac{\delta^{(D)}(x-z)}{\sqrt{-g}} ,
\end{equation}
\end{subequations}
where $B^{\mu\nu}(\xi)$ and $C^{\lambda\mu\nu}(\xi)$ are the corresponding
multipole coefficients. The de\-com\-po\-si\-ti\-on (\ref{jna5})
is used as an ansatz for solving the conservation equations
(\ref{jna3}). This has already been done in \cite{12a}, resulting in
manifestly covariant $p$-brane world-sheet equations.

In the next section, the general result of \cite{12a} is applied to the
$p=2$ case. In particular, the Nambu-Goto type of membrane with totally
antisymmetric spin tensor parallel to the world-sheet is thoroughly
examined.

\section{Membrane dynamics}\label{Sec3}

In this section, the stress-energy and spin tensor conservation equations
(\ref{jna3}) are solved in the single-pole approximation. The general brane
world-sheet equations and boundary conditions are applied to the membrane
case.

\subsection{Preliminaries}

The $p$-brane world-sheet equations in the single-pole approximation are
obtained in the following way. We insert the ansatz (\ref{jna5})
into the conservation equations (\ref{jna3}), and solve for
the unknown variables $z^{\mu}(\xi)$, $B^{\mu\nu}(\xi)$ and
$C^{\lambda\mu\nu}(\xi)$. The algorithm for solving this type of equation
has been discussed in detail in \cite{12,12a}, and here we use the
ready-made result. According to \cite{12a}, the single-pole world-sheet
equations are given by
\begin{subequations} \label{jna6}
\begin{equation} \label{jna6a}
\Pn^{\mu}_{\rho} \Pn^{\nu}_{\sigma} D^{\rho\sigma} =0,
\end{equation}
\begin{equation} \label{jna6b}
\begin{array}{c}
\ds\nabla_a\left( m^{ab}u_b^{\mu} - 2 u^a_{\rho} D^{\mu\rho} + u_b^{\mu}
u^b_{\rho} u^a_{\sigma} D^{\rho\sigma} \right) = \\
\ds =\frac{1}{2}C_{\nu\rho\sigma} \nabla^{\mu} K^{\rho\sigma\nu},
\end{array}
\end{equation}
\end{subequations}
while the boundary conditions have the form
\begin{equation} \label{jna7}
n_a \left( m^{ab}u_b^{\mu} - 2 u^a_{\rho} D^{\mu\rho} + u_b^{\mu} u^b_{\rho}
u^a_{\sigma} D^{\rho\sigma} \right) \Big|_{\del\cM}=0.
\end{equation}
Here, $\Pn^{\mu}_{\nu} \equiv \delta^{\mu}_{\nu} - u_a^{\mu}u^a_{\nu}$ is the
orthogonal world-sheet projector, $n^a$ is the unit boundary normal, and
$\nabla_a$ stands for the total covariant derivative that acts on both types of
indices:
$$
\nabla_a V^{\mu b} \equiv  \del_a V^{\mu b} + \cfl{\mu}{\lambda\rho}
u_a^{\rho} V^{\lambda b} + \cfl{b}{ca}V^{\mu c}.
$$
The coefficients $m^{ab}(\xi)$ and $C^{\lambda\mu\nu}(\xi)$ are the residual
free parameters of the theory. While $m^{ab}$ represents the effective
stress-energy tensor of the brane, the $C^{\lambda\mu\nu}$ currents are
related to its spin density. The shorthand notation
$$
D^{\mu\nu} \equiv K^{[\mu}{}_{\lambda\rho} C^{\rho\lambda\nu]} + \frac{1}{2}
K_{\lambda\rho}{}^{[\mu} C^{\nu]\rho\lambda}
$$
is introduced for convenience.

The world-sheet equations (\ref{jna6}) and boundary
conditions (\ref{jna7}) describe the dynamics of an infinitely thin
$p$-brane in $D$-dimensional spacetime with curvature and torsion. We want
to emphasize that the obtained dynamics has rather universal character. This
is because our derivation rests upon the mere existence of the conservation
equations of the stress-energy and spin tensors. These conservation
equations are the direct consequence of the diffeomorphism and local Lorentz
symmetry of the matter Lagrangian. As a consequence of the universality of
these symmetries, the conservation equations hold regardless of the details
of the theory. This is the virtue of our approach, as it is independent of
the particular action used. On the other hand, the conservation equations
have the weakness of being incomplete, in the sense that they carry much
less information than the full set of field equations. As a consequence, the
derived $p$-brane world-sheet equations contain free coefficients. The
details of the initial theory are hidden in the form of these coefficients.

By inspecting the form of the obtained world-sheet equations, we realize
that only branes made of spinning matter can probe spacetime torsion.
Moreover, if the torsion is totally antisymmetric, only axial component of
the spin tensor $C^{\lambda\mu\nu}$ survives in the world-sheet equations.
In this paper, we shall examine how axial spin tensors couple to torsion. In
particular, we are interested in branes characterized by maximally symmetric
distribution of stress-energy and spin. It has already been shown in
\cite{12a} that this task can not be accomplished with strings. The reason
for this is that the only candidate for the right type of spin-torsion
coupling turns out to be the projection of the axial $C^{\lambda\mu\nu}$ on
the world-sheet, and it identically vanishes if $p=1$. In what follows, we
shall turn to membranes ($p=2$), and demonstrate how the choice of axial
$C^{\lambda\mu\nu}$ parallel to the world-sheet solves the problem.

\subsection{World-sheet equations}

Let us consider an infinitely thin membrane whose spin tensor is totally
antisymmetric and parallel to the world-sheet. It is defined by the
$C^{\lambda\mu\nu}$ coefficient of the form
\begin{equation} \label{jna8}
C^{\lambda\mu\nu} = s e^{abc} u_a^{\lambda} u_b^{\mu} u_c^{\nu}\,,
\end{equation}
where $e^{abc}$ is the covariant Levi-Civita symbol, and $s$ is a scalar
which measures the spin magnitude. In addition, we shall restrict the scalar
$s$ to be a constant, so that our membrane has maximally symmetric
distribution of spin. Indeed, the constant $s$ is the only variable in
(\ref{jna8}) which carries the information about the nature of matter fields
the membrane is made of. The rest of $C^{\lambda\mu\nu}$ is given in pure
geometric terms. At any given point of the world-sheet, these terms can be
gauged away by the proper choice of the target space and world-sheet
coordinates. Precisely, we can achieve $g_{\mu\nu}=\eta_{\mu\nu}$ and
$u_a^{\mu}=\delta_a^{\mu}$, so that the non-vanishing components of
$C^{\lambda\mu\nu}$ reduce to $C^{abc}=s\epsilon^{abc}$. The Levi-Civita
symbol $\epsilon^{abc}$ is a constant tensor of the Lorentz group, which
ensures local isotropy of $C^{\lambda\mu\nu}$. As the whole procedure can be
repeated with a different choice of the word-sheet point, the
$C^{\lambda\mu\nu}$ is a homogeneous tensor, as well. This is exactly what
ensures maximal symmetry of the spin tensor.

It is immediately seen that this kind of spin tensor does not exist in the
string case. There, the world-sheet is $2$-dimensional, and the
corresponding Levi-Civita tensor is a second rank tensor. In the membrane
case, the world-sheet indices $a,b,\dots$ take three values, exactly as
needed for the existence of the third rank Levi-Civita tensor. As we have
seen, this tensor is indispensable for the construction of the maximally
symmetric spin tensor.

Let us now see how the membrane spin tensor of the form (\ref{jna8})
influences the world-sheet equations (\ref{jna6}). First,
we calculate the $D^{\mu\nu}$ tensor, and find that it reduces to
\begin{equation} \label{jna9}
D^{\mu\nu} = \frac{3s}{2} e_{abc}u^{c[\mu} \Ka^{ab\nu]} .
\end{equation}
It depends on the axial part of the contorsion
$$
\Ka^{\mu\nu\rho} \equiv \frac{1}{3} \left( K^{\mu\nu\rho} + K^{\nu\rho\mu} +
K^{\rho\mu\nu}  \right)
$$
through its world-sheet projection $\Ka^{ab\rho} \equiv u^a_{\mu} u^b_{\nu}
\Ka^{\mu\nu\rho}$. The precession equations (\ref{jna6a}) are identically
satisfied, and we are left with the world-sheet equations (\ref{jna6b}) and
boundary conditions (\ref{jna7}). Now, we calculate the right-hand side of
(\ref{jna6b}), and find
\begin{equation} \label{jna10}
C_{\nu\lambda\rho} \nabla^{\mu} K^{\nu\lambda\rho} = s u_{\nu\lambda\rho}
\Ka^{\mu\nu\lambda\rho} + \nabla^c \left( 3s e_{abc} \Ka^{ab\mu} \right),
\end{equation}
with
$$
\Ka^{\mu\nu\lambda\rho} \equiv \nabla^{\mu} \Ka^{\nu\lambda\rho} -
\nabla^{\nu} \Ka^{\lambda\rho\mu} + \nabla^{\lambda} \Ka^{\rho\mu\nu} -
\nabla^{\rho} \Ka^{\mu\nu\lambda},
$$
and
$$
u^{\mu\nu\rho} \equiv e^{abc} u_a^{\mu} u_b^{\nu} u_c^{\rho} .
$$
With the help of (\ref{jna9}) and (\ref{jna10}), the world-sheet equations
are rewritten as
\begin{equation} \label{jna11}
\nabla_b \left( \tm^{ab} u_a^{\mu} \right) = \frac{s}{2} u_{\nu\lambda\rho}
\Ka^{\mu\nu\lambda\rho},
\end{equation}
where $\tm^{ab} \equiv m^{ab} - \frac{s}{2} \gamma^{ab} K^{\mu\nu\rho}
u_{\mu\nu\rho}$ are the residual free coefficients. Using the total
antisymmetry of the $\Ka^{\mu\nu\lambda\rho}$ field in (\ref{jna11}), the
coefficients $\tm^{ab}$ are shown to be covariantly conserved,
\begin{equation} \label{jna12}
\nabla_b \tm^{ab} =0.
\end{equation}
This means that Nambu-Goto matter is allowed as the constituent matter of
our membrane. Indeed, by demanding
\begin{equation} \label{jna13}
\tm^{ab} = T \gamma^{ab},
\end{equation}
where $T$ is a constant commonly interpreted as the membrane tension, the
condition (\ref{jna12}) is automatically satisfied. At the same time, this
choice ensures maximally symmetric distribution of the membrane
stress-energy. That it is indeed so can be seen by repeating the arguments
used for the demonstration of maximal symmetry of the spin tensor.

Now we are left with only two constants, $T$ and $s$, to characterize our
membrane. With this, the world-sheet equations receive their final form
\begin{equation} \label{jna14}
\nabla_a u^{a\mu} = \frac{s}{2T} \Ka^{\mu\nu\lambda\rho} u_{\nu\lambda\rho} .
\end{equation}
Following the same procedure, the boundary conditions are rewritten as
\begin{equation} \label{jna15}
n^c \left( u_c^{\mu} + \frac{3s}{2T} e_{abc} \Ka^{ab\mu} \right)
\Big|_{\del\cM} =0.
\end{equation}
The world-sheet equations (\ref{jna14}), and the boundary conditions
(\ref{jna15}) govern the dynamics of a Nambu-Goto type of membrane with
axial spin tensor parallel to the world-sheet. Such a membrane is
characterized by two constants, the tension $T$ and the spin magnitude $s$,
and represents a minimal extension of the Nambu-Goto case.

The equations (\ref{jna14}) and (\ref{jna15}) are shown to follow from an
action functional. Indeed, they are obtained by varying the action
\begin{equation} \label{jna16}
\begin{array}{ccl}
I & = & \ds T\int d^3\xi\sqrt{-h}\Big[g_{\mu\nu}(x)
u_a^{\mu} u_b^{\nu} h^{ab} +                              \\
& & \ds \hphantom{T\int d^3\xi\sqrt{-h}} +
\frac{s}{T} K_{\mu\nu\rho}(x)u_a^{\mu} u_b^{\nu}
u_c^{\rho} e^{abc}-1\Big]                                 \\
\end{array}
\end{equation}
with respect to the independent variables $x^{\mu}(\xi)$ and $h_{ab}(\xi)$.
We cannot help noticing the resemblance of this action to the $\sigma$-model
action of \cite{22}. The latter describes an elementary membrane interacting
with the 3-form field $\cB_{\mu\nu\rho}$. In fact, the two actions differ in
one instance only: the role of the 3-form field $\cB_{\mu\nu\rho}$ in
\cite{22} is played by the contorsion field $\frac{s}{T}\Ka_{\mu\nu\rho}$ in
(\ref{jna16}). Although this matching may be just a coincidence, we find it
appropriate to draw the reader's attention to it.

In the next section, we shall demonstrate how the effective string dynamics
is obtained in the narrow membrane limit. This way, the dimensionally
reduced analogue of the action functional (\ref{jna16}) will be obtained.

\section{Dimensional reduction}\label{Sec4}

The results of the preceding section are obtained under very general
assumptions concerning the dimensionality and topology of spacetime and
world-sheet. In what follows, we shall use this freedom to apply these
results to a cylindrical membrane wrapped around the extra compact dimension
of a $(D+1)$-dimensional spacetime. In the limit of a narrow membrane, we
expect to obtain the effective string dynamics. In fact, this kind of double
dimensional reduction has already been considered in \cite{22}. There, the
string effective action in $10$ dimensions has been obtained from the
membrane action in $11$ dimensions. To complete our exposition, we shall
describe a similar $D+1\to D$ dimensional reduction.

Let us consider a $(D+1)$-dimensional spacetime with one small compact
dimension. It is parametrized by the coordinates $X^M$ ($M=0,1,\dots,D$),
which we divide into the ``observable'' coordinates $x^{\mu}$ ($\mu =
0,1,\dots,D-1$), and the extra periodic coordinate $y$. In the limit of
small extra dimension, we use the Kaluca-Klein ansatz
\begin{equation} \label{jna18}
\del_y K_{MNL} =0, \qquad \del_y G_{MN}=0, \qquad G_{yy}=1
\end{equation}
to model the contorsion and metric. This ansatz violates the
$(D+1)$-dimensional diffeomorphisms, leaving us with the coordinate
transformations
$$
x^{\mu'} = x^{\mu'} (x), \qquad y' = y + \lc(x).
$$
In what follows, we shall use the decomposition
\begin{equation} \label{jna19}
G_{MN} = \left(
\begin{array}{cc}
g_{\mu\nu} + a_{\mu}a_{\nu} & a_{\mu} \\
a_{\nu} & 1 \\
\end{array} \right) ,
\end{equation}
as it yields the variables that transform as tensors with respect to the
residual $D$-diff\-e\-o\-mor\-ph\-i\-sms. The same kind of argument applies
to $K_{MNL}$. We shall use a \textit{totally antisymmetric contorsion}, and
decompose it as
\begin{equation} \label{jna20}
K^{MN}{}_L \equiv
\Big( K^{\mu\nu}{}_{\lambda}\,,\, K^{\mu\nu}{}_y \Big) \equiv
\Big( \cK^{\mu\nu}{}_{\lambda} +
k^{\mu\nu}a_{\lambda}\,,\, k^{\mu\nu} \Big).
\end{equation}
The components $\cK^{\mu\nu}{}_{\lambda}$ and $k^{\mu\nu}$ are tensors with
respect to the residual diffeomorphisms. Now, we consider a membrane wrapped
around the extra compact dimension $y$. Its world-sheet $X^M=Z^M(\xi^A)$ is
denoted by $\cM_3$, and is chosen in the form
\begin{equation} \label{jna21}
x^{\mu} = z^{\mu}(\xi^a), \qquad y=\xi^2,
\end{equation}
where the world-sheet coordinates $\xi^A$ ($A=0,1,2$) are divided into
$\xi^a$ ($a=0,1$) and $\xi^2$. This ansatz reduces the reparametrizations
$\xi^{A'}= \xi^{A'}(\xi^B)$ to
$$
\xi^{a'} = \xi^{a'}(\xi^b), \qquad \xi^{2'} = \xi^2 + \lc(z^{\mu}(\xi)),
$$
and the world-sheet tangent vectors $U_A^M = \del Z^M / \del\xi^A$ to
\begin{equation} \label{jna22}
U_a^{\mu} = u_a^{\mu}, \qquad U_2^{\mu} = U_a^y =0, \qquad U_2^y=1.
\end{equation}
One can verify that $u_a^{\mu} \equiv \del z^{\mu} / \del\xi^a$ transforms
as a tensor with respect to the residual spacetime and world-sheet
diffeomorphisms. The induced metric $\itGamma_{AB} \equiv G_{MN} U_A^M
U_B^N$ is shown to satisfy the condition $\del_2 \itGamma_{AB} =0$.
It is decomposed as
\begin{equation} \label{jna23}
\itGamma_{AB} = \left(
\begin{array}{cc}
\gamma_{ab} + a_aa_b & a_a \\
a_b & 1 \\
\end{array} \right) ,
\end{equation}
with $\gamma_{ab}\equiv g_{\mu\nu} u_a^{\mu} u_b^{\nu}$, and $a_a \equiv
a_{\mu} u_a^{\mu}$. In what follows, we shall refer to
$x^{\mu}=z^{\mu}(\xi^a)$ as the string world-sheet, and denote it by
$\cM_2$.

The membrane boundary $\del\cM_3$ is given by $\xi^A = \zeta^A(\lambda^i)$,
where $\lambda^i$ ($i=0,1$) are the boundary coordinates. In accordance with
the ansatz (\ref{jna21}), it is chosen in the form
\begin{equation} \label{jna24}
\xi^a = \zeta^a(\lambda^0), \qquad \xi^2 = \lambda^1 .
\end{equation}
The boundary tangent vectors $V^A_i \equiv \del \zeta^A / \del\lambda^i$ and
the boundary normal $N_A \equiv e_{ABC}V_0^B V_1^C$ are thereby reduced to
$v^a=\del \zeta^a / \del\lambda^0$ and $n_a = e_{ab} v^b$, respectively. In
what follows, we shall refer to $\xi^a=\zeta^a(\lambda^0)$ as the string
boundary $\del\cM_2$.

The membrane world-sheet equations of section \ref{Sec3} are now rewritten
as
\begin{equation} \label{jna25}
\nabla_A U^{AM} = \frac{s}{2T} e_{ABC} U^A_N U^B_L U^C_R K^{MNLR},
\end{equation}
with
$$
\begin{array}{ccl}
K^{MNLR} & \equiv & \nabla^M K^{NLR} - \nabla^N K^{LRM} + \\
& & + \nabla^L K^{RMN} - \nabla^R K^{MNL}. \\
\end{array}
$$
Using the Kaluca-Klein ansatz (\ref{jna18}), (\ref{jna21}), and the
decompositions (\ref{jna19}), (\ref{jna20}), (\ref{jna22}) and
(\ref{jna23}), one finds the following. The $M=y$ component of the
world-sheet equations (\ref{jna25}) is identically satisfied. The $M=\mu$
components become
\begin{equation} \label{jna26}
\nabla_a u^{a\mu} = \frac{3s}{2T} e_{ab} u^a_{\nu}u^b_{\lambda}
k^{\mu\nu\lambda},
\end{equation}
with
$$
k^{\mu\nu\lambda} \equiv \nabla^{\mu} k^{\nu\lambda} + \nabla^{\nu}
k^{\lambda\mu} + \nabla^{\lambda} k^{\mu\nu}.
$$
Similarly, the $M=y$ component of the boundary conditions
\begin{equation} \label{jna27}
N^C \left( U_C^M + \frac{3s}{2T} e_{ABC} K^{ABM} \right) \Big|_{\del\cM_3}=0
\end{equation}
is identically satisfied, while $M=\mu$ components reduce to
\begin{equation} \label{jna28}
n^a \left( u_a^{\mu} + \frac{3s}{T} e_{ab} k^{\mu b} \right)
\Big|_{\del\cM_2} =0 .
\end{equation}
The world-sheet equations (\ref{jna26}) and boundary conditions
(\ref{jna28}) are shown to follow from the action functional
\begin{equation} \label{jna29}
I=T \int \rmd^2\xi \sqrt{-h} \left[ g_{\mu\nu}(x) h^{ab} +
\frac{3s}{T}k_{\mu\nu}(x)e^{ab} \right] u_a^{\mu} u_b^{\nu}\,.
\end{equation}
Again, we notice a similarity between our action (\ref{jna29}) and the
string sigma model action of \cite{15,16,17,18,t7,t19}. The latter describes
an elementary string coupled to the string axion $\cB_{\mu\nu}$. As in the
membrane case, the two actions differ in one instance only: the role of the
string axion $\cB_{\mu\nu}$ in \cite{15,16,17,18,t7,t19} is played by the
extra dimensional components of the contorsion $\frac{3s}{T} k_{\mu\nu}$ in
(\ref{jna29}). Whether this is just a coincidence, or there is more content
in this matching is not the subject of our paper. Anyway, it is an
interesting observation which, we believe, deserves to be mentioned.

In summary, we have derived how uniform membranes made of spinning matter
behave in Riemann-Cartan backgrounds. We have considered a minimal extension
of the Nambu-Goto membrane, characterized by two constants only: the tension
and spin magnitude. Such membranes proved to be the simplest branes with a
nontrivial spin-torsion coupling. The effective string dynamics is obtained
from cylindrical membranes in the narrow membrane limit. Both sets of
equations are found to follow from corresponding action functionals. By
inspecting their form we have discovered a similarity with string theory
sigma models.

\section{Concluding remarks}\label{Sec5}

In this paper, we have analyzed the behaviour of classical membranes in
Riemann-Cartan backgrounds. The membrane constituent matter is specified in
terms of its stress-energy and spin tensors. In particular, the
stress-energy is assumed to be of the Nambu-Goto type, and the spin tensor
is chosen totally antisymmetric and parallel to the world-sheet. Such
membranes have maximally symmetric distribution of stress-energy and spin,
and are characterized by two constants only: the tension and spin magnitude.
The idea behind these considerations is the search for the simplest brane
with nontrivial spin-torsion coupling.

The method we use is a generalization of the Math\-is\-s\-on-Papapetrou
method for pointlike matter \cite{1,2}. It has already been used in
\cite{11,12,12a} for the study of strings and higher branes in Riemann-Cartan
backgrounds. In this work, the general results of \cite{12a} have been
applied to a membrane with maximally symmetric distribution of stress-energy
and spin. The effective string is seen as a narrow membrane wrapped around
the extra spatial dimension.

Our exposition is summarized as follows. In section \ref{Sec2}, we have
reviewed the basics of the multipole formalism developed in \cite{11,12,12a}.
A manifestly covariant multipole expansion has been defined for an arbitrary
exponentially decreasing function, and then applied to the stress-energy and
spin tensors of localized matter. The dynamics is specified through the
stress-energy and spin tensor covariant conservation equations.

In section \ref{Sec3}, the conservation equations have been solved for an
arbitrary infinitely thin $p$-brane. The resulting manifestly covariant
world-sheet equations and boundary conditions have then been applied to the
$p=2$ case. The motivation comes from the observation that the needed
world-sheet projection of the axial spin tensor vanishes if $p=1$. This kind
of spin tensor proved to be the basic ingredient for the construction of a
membrane with maximally symmetric distribution of spin. The obtained
membrane dynamics has been verified to follow from an action functional. A
resemblance with the $\sigma$-model action of \cite{22} has been noted.

In section \ref{Sec4}, we have considered cylindrical membranes wrapped
around the extra compact dimension of a $(D+1)$-dimensional spacetime. The
effective string dynamics is obtained in the narrow membrane limit. It turns
out to resemble the string dynamics of \cite{15,16,17,18,t7,t19}. Precisely,
we have noticed that macroscopic strings couple to torsion the same way as
fundamental strings couple to the string axion.

In summary, let us say something about the prospects of our research. We are
aware of the fact that considerable additional work can be done along the
lines followed in this paper. In fact, our world-sheet equations
(\ref{jna6}) and boundary conditions (\ref{jna7}) contain free parameters
that can be chosen in a variety of ways, each leading to a different brane
dynamics. The simple choice considered in this paper (membranes with
constant tension and spin density), has been shown to lead to the familiar
membrane dynamics, and after simple compactification, to the familiar string
dynamics. An interesting possibility is to consider more general membranes,
and more general compactifications. Hopefully, couplings to the
electromagnetic and scalar fields could be discovered in the dimensionally
reduced theory. This is, however, a difficult task for itself, and will be
considered in a separate paper.

\begin{acknowledgments}
This work is supported by the Serbian Ministry of Science and Technological
Development, under Contract No. $141036$.
\end{acknowledgments}

\end{document}